\documentclass[sigconf]{acmart}

\usepackage{verbatim}
\usepackage{tabularx}
\usepackage{balance}

\usepackage{array}
\usepackage{graphicx}
\usepackage{todonotes}
\usepackage{xcolor}
\usepackage{subcaption}
\usepackage{wrapfig}
\usepackage{arydshln}
\usepackage{ulem}
\usepackage{textgreek}
\usepackage{pgfplots}
\usepackage{commath}
\usepackage{multirow}
\usepackage{makecell}
\usepackage[most, breakable]{tcolorbox}
\usepackage{times}
\usetikzlibrary{calc}
\pgfplotsset{compat=1.12}

\definecolor{darkgreen}{RGB}{21,176,26}
\definecolor{cRed}{rgb}{0.648, 0.109, 0}
\definecolor{cGreen}{rgb}{0.0, 0.496, 0.05}
\definecolor{hypothesis-color}{RGB}{136, 182, 209}

\newcolumntype{P}[1]{>{\centering\arraybackslash}p{#1}}
\newcolumntype{M}[1]{>{\centering\arraybackslash}m{#1}}
\newcolumntype{N}{@{}m{0pt}@{}}

\graphicspath{{figures/}}

\definecolor{bleudefrance}{rgb}{0.19, 0.55, 0.91}

\begin{document}
	
\title[Lifelong Computing]{Lifelong Computing}

\author{Danny Weyns}
\affiliation{
	\institution{KU Leuven, Belgium}
	\country{Linnaeus University, Sweden}
}
\email{danny.weyns@kuleuven.be}

\author{Thomas B\"ack}
\orcid{0000-0001-6768-1478}
\affiliation{
	\institution{Leiden University, The Netherlands}
	\country{NORCE Norwegian Research Centre}
}
\email{t.h.w.baeck@liacs.leidenuniv.nl}

\author{Ren\'e Vidal}
\affiliation{
	\institution{Johns Hopkins University, USA}
	\country{NORCE Norwegian Research Centre}
}
\email{rvidal@jhu.edu}

\author{Xin Yao}
\affiliation{
	\institution{University of Birmingham, UK}
	\country{SUSTech, China}
}
\email{xiny@sustech.edu.cn}

\author{Ahmed Nabil Belbachir}
\affiliation{
	\institution{NORCE Norwegian Research Centre}
	\country{}
}
\email{nabe@norceresearch.no}

\begin{abstract}

Computing systems form the backbone of many aspects of our life, hence they are becoming as vital as water, electricity, and road infrastructures for our society. Yet, engineering long running computing systems that achieve their goals in ever-changing environments pose significant challenges. Currently, we can build computing systems that adjust or learn over time to match changes that were anticipated. However, dealing with unanticipated changes, such as anomalies, novelties, new goals or constraints, requires system evolution, which remains in essence a human-driven activity. Given the growing complexity of computing systems and the vast amount of highly complex data to process, this approach will eventually become unmanageable. To break through the status quo, we put forward a new paradigm for the design and operation of computing systems that we coin \textit{lifelong computing}. The paradigm starts from computing-learning systems that integrate computing/service modules and learning modules. Computing warehouses offer such computing elements together with data sheets and usage guides. When detecting anomalies, novelties,  new goals or constraints, a lifelong computing system activates an evolutionary self-learning engine that runs online experiments to determine how the computing-learning system needs to evolve to deal with the changes, thereby changing its architecture and integrating new computing elements from computing warehouses as needed. Depending on the domain at hand, some activities of lifelong computing systems can be supported by humans. We motivate the need for lifelong computing with a future fish farming scenario, outline a blueprint architecture for lifelong computing systems, and highlight key research challenges to realise the vision of lifelong computing. 

\end{abstract}


\maketitle

\section{Introduction}

In the emerging hyper-connected digital world, computing systems will need to work radically different than today, ubiquitously connected to a tremendous amount of diverse data and facing a priori unknown conditions during operation. This digitisation penetrates every aspect of our life, from healthcare and industries, to traffic, telecommunication, finance, and environment security. As such, computing systems are becoming as vital as water, electricity, and road infrastructures for our society.
Consequently, we increasingly depend on the trustworthiness and sustainability of computing systems~\cite{EU,NSF}. Yet, engineering these systems is extremely challenging due to complexity that arises from the ever changing conditions they face, such as dynamics in the environment, new emerging technologies, and the need to deal with new goals or constraints. 

Our current knowledge allows building computing systems that can deal with changes that were anticipated. To that end, some of the tasks can be automated, for instance, systems can adjust themselves or learn over time to match changes in their environment, or tools can be used to automate system updates. Other tasks can be managed by system operators, e.g., start/stop tasks, or perform maintenance. 
However, we lack knowledge about how to build computing systems that can deal with unanticipated changes autonomously. 

\begin{figure*}[!thb]
    \centering
    \includegraphics[width=0.82\linewidth]{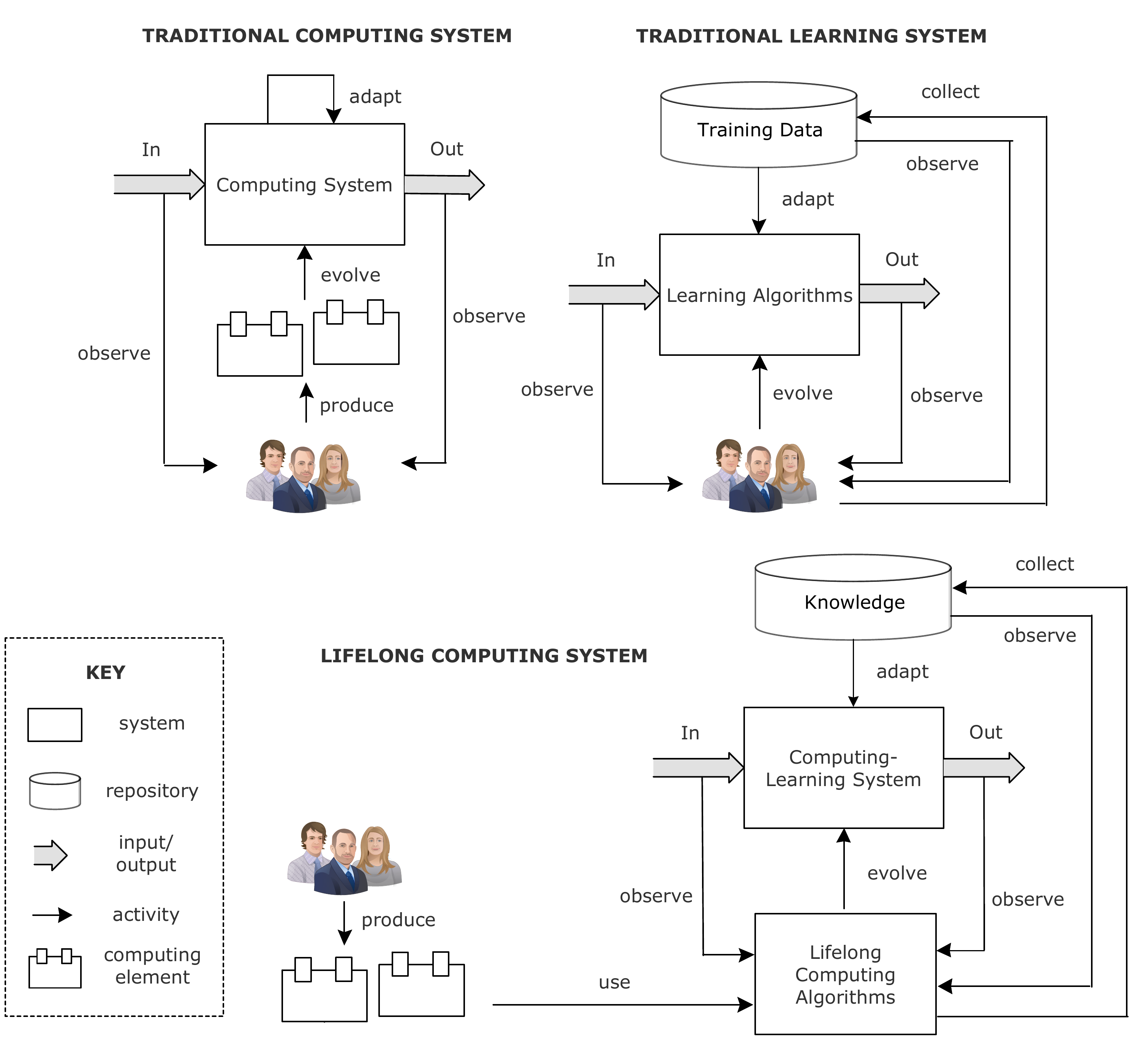}
    \caption{From traditional computing and learning systems to lifelong computing systems. 
    }
    \label{fig:overview}
\end{figure*}

Dealing with anomalies, novel conditions, new goals or constraints typically requires system evolution, which remains in essence a human-driven activity. With the ever increasing complexity of computing systems and the vast amount of highly complex data these systems need to process, human-driven approaches will eventually become unmanageable. The availability of sensors, the capacity to handle huge amounts of data, and the processing capacity and methods to run the necessary decision algorithms opens perspectives to major breakthroughs towards fully autonomous systems that operate in complex environments~\cite{DNV}. However, we currently lack fundamental knowledge to turn these long-standing and emerging challenges into reality. 
To break through the status quo, we put forward a new paradigm for designing and operating computing systems that we coin \textit{lifelong computing}. 
As illustrated in Figure~\ref{fig:overview}, lifelong computing systems differ fundamentally from traditional computing systems and learning algorithms by using lifelong \mbox{computing algorithms in lieu of humans to adapt and evolve.}

A traditional computing system takes inputs from the environment and produces outputs realising users' goals.\footnote{We use goals, requirements, and objectives interchangeably in this paper.} 
To deal with changing conditions, such a system can be equipped with artificial intelligence (AI) techniques enabling it to \textit{operate autonomously} when facing changes, or alternatively by enhancing it with a feedback loop, enabling the system to \textit{self-adapt} its configuration autonomously to deal with changes. 
However, traditional computing systems are designed to work in an operational domain; dealing with unanticipated changes requires in general an evolution of the system. \textit{Software evolution} typically relies on humans that produce new computing elements. These elements are then integrated in the system, a process that is more and more automated. To find near-optimal solutions, software engineers increasingly make use of \textit{search-based techniques} that rely on meta-heuristic search methods, such as genetic algorithms. These techniques enable modifying a software component to make it more efficient in terms of performance and resource use. 

A traditional learning algorithm on the other hand is  trained to perform a set of pre-defined tasks using data that are often labelled by humans. When such a system is deployed, the distribution of the data might be different, and additional training data might be required to fine-tune the learning algorithm to new conditions. This evolution step typically relies on human intervention. \textit{Anomaly and novelty detection} mechanisms can be used to identify data points that deviate from a data set's normal behaviour. \textit{Lifelong learning} aims at automating the evolution of a learning algorithm by adding a meta-learner on top of the algorithm enabling it to deal with new tasks. However, lifelong learning 
algorithms are still in early stages of development and far from being deployed in real-life settings.

In contrast, a lifelong computing system maintains knowledge about itself and its goals (self-awareness), and its environment (context-awareness). The lifelong computing algorithms use this knowledge to autonomously self-adapt and self-evolve the architecture of the system, dealing with anticipated and unanticipated changes respectively, throughout the system's lifetime. 
To provide broad functionalities and handle vast amounts of complex data, these systems are becoming increasingly heterogeneous, integrating computing elements with learning algorithms. For their evolution lifelong computing systems can autonomously integrate new computing elements or learning algorithms produced by humans. Lifelong computing resembles the idea of  ``self-growing software'' proposed by Tamai~\cite{Tamai2019} as the next paradigm shift in software engineering.  

The remainder of this paper starts with highlighting current approaches to deal with change and arguing  why a novel foundation is required.
Then we outline a blueprint architecture for lifelong computing systems.
We use a future offshore fish farming scenario as an illustrative example. 
To conclude, we summarise the novelty of lifelong learning, and from that, we outline key research challenges associated with realising the vision of lifelong computing. 

\section{State of Affairs}\label{sec:soa}

Dealing with change has been a major aspect of computing systems since their inception. We discuss six main lines of work that have been studying how to deal with changes as highlighted above: autonomous systems, self-adaptation, software evolution, search-based software engineering, anomaly and novelty detection, and lifelong learning. We then argue why a novel holistic approach is required to deal with the challenges of future computing systems. 

\vspace{5pt}\noindent\textbf{Autonomous Systems}. 
Autonomous systems (or intelligent autonomous systems) are computing systems that act independently of direct human supervision~\cite{Tzafestas2012}. A central idea of autonomous systems is to mimic human (or animal) 
intelligence, which has been a source of inspiration for a very long time. Wiener, who invented cybernetics at MIT in the 1950s, laid the foundations of fields such as feedback control, automation, and robotics. Autonomous systems exploit AI techniques for instance for sensing and perception, data processing and information fusion, intelligent decision making, and interaction and cooperation. Important sub-fields of autonomous systems are multi-agent systems~\cite{Wooldrige2009} that studies the coordination of autonomous agents to solve problems that go beyond the capabilities of single agents, and human-robot teams~\cite{MUSIC201642} that studies collaboration of humans and robots exploiting their complementary skill sets. The interest in autonomous systems expanded significantly in recent years, with high-profile applications such as self-driving cars, smart manufacturing robots (Industry 4.0 driven by the Internet of Things), and care robots for the elderly. While having extreme potential, the real technical difficulties associated with realising truly autonomous systems remain a key challenge as illustrated by examples of accidents caused by first generation autonomous cars~\cite{7573335}. 

\vspace{5pt}\noindent\textbf{Self-Adaptation}. Back in 2000, IBM released a manifesto that referred to a ``looming software crisis'' caused by the increasing complexity of installing, configuring, tuning, and maintaining computing systems~\cite{Kephart}. A consensus grew that self-adaptation\footnote{The term self-adaptation is also used in evolutionary computation, where it denotes approaches for continuously adapting the sampling distribution during the search process, based on the history of evaluated solution candidates~\cite{Baeck1998,Meyer-Nieberg2007}.} was the only viable option to tackle the problems that caused this complexity crisis. Since then, extensive efforts have been put in devising fundamental principles of self-adaptation as well as techniques and methods to engineer self-adaptive systems~\cite{weyns2020book}. Over time, the focus shifted from automating operator tasks based on high-level goals~\cite{Kephart,Rainbow} to taming uncertainties that computing systems face during operation and that are difficult to anticipate before deployment~\cite{2786805.2786853,8008800}. At the heart of self-adaptive systems are runtime models~\cite{Blair2009} that provide the system with self-awareness (self-representation and representation of goals) and context-awareness (representation of the environment)~\cite{10.1145-3347269,weyns2020book}. These models are updated at runtime and used to analyse the situation and decide when and how to adapt the system to maintain its goals. Open challenges include principled solutions to self-adaptation of large-scale systems, dealing with unanticipated changes, the exploitation of AI techniques in the realisation of self-adaptation, and establishing trust in self-adaptive systems~\cite{Cheng2009,weyns2020book}. 

\vspace{5pt}\noindent\textbf{Software Evolution}. 
During the past decades, the traditional view of software that evolves through periodic releases has been replaced by continuous evolution of software~\cite{RODRIGUEZ2017263}. Software organisations today develop, release and learn from software in rapid parallel cycles (from hours to a few weeks). This approach is commonly referred as continuous deployment (CD)~\cite{978-3-319-08738}. CD is based on the principles of agile development~\cite{DINGSOYR20121213} and DevOps~\cite{MISHRA2020100308} that aim at increasing the deployment speed and quality. CD leverages on continuous integration (CI)~\cite{6802994} that automates tasks such as compiling code, running tests, and building deployment packages. Among the benefits of CI/CD are rapid innovation, shorter time-to-market, increased customer satisfaction, continuous feedback, and improved developer productivity. Yet, an important concern of current practice in software maintenance is (intentional or unintentional) technical debt, i.e., 
longer-term negative effects on systems that result from sub-optimal decisions~\cite{LI2015193} in agile development. Key research challenges in software evolution include mechanisms to (automatically) add/exchange functionality, continuous experimentation, continuous quality assessment, and effective exploitation of user feedback~\cite{RODRIGUEZ2017263}. 

\vspace{5pt}\noindent\textbf{Search Based Software Engineering (SBSE)}. The idea of SBSE is to formulate a software engineering problem as an optimisation problem and then apply a meta-heuristic search method to solve that problem~\cite{Harman2001}. Compared to traditional exact techniques, such as linear and dynamic programming, meta-heuristic search methods can handle complex representations of search domains and scale much better. Such search methods have been used to find the best subset of requirements that balance cost, risk, and user satisfaction~\cite{7582553}, automatic generation of test cases~\cite{7102580}, to apply genetic techniques for repairing programs by altering a few lines of source code~\cite{6227211}, and to embed optimisation into the deployed software to create self-optimising systems~\cite{6475391}. Besides challenges related to its practical applicability (e.g., non-determinism of results), meta-heuristic search methods are subject to other challenges related to its automation, such as dividing the line between adaptive automation for small changes and human intervention to invoke more fundamental adaptation, and efficient 
(runtime) computation of large numbers of fitness evaluations between adaptations~\cite{10.1145-3449726.3461425}. 

\vspace{5pt}\noindent\textbf{Anomaly and Novelty Detection}. %
Anomaly and novelty detection (or outlier detection) aims at recognising data instances that significantly deviate from the majority of data instances~\cite{1969Grubbs}. It has been used in a variety of domains, e.g., intrusion detection, fault prevention, defect detection, and unexpected flow  detection. A plethora of methods have been developed~\cite{3381028}, including proximity-based approaches that rely on relations between nearby data points, projection techniques that convert data into a space with reduced dimensionality to improve outlier detection, outlier detection for high-dimensional data such as recursive binning and re-projection, windowing for online time series that incrementally builds and updates models with new data, and deep learning anomaly detection, such as deep neural network auto-encoders. Challenges include novelty detection on noisy high-dimensional data and over data streams, detection of complex anomalies, and interpretable deep anomaly detection~\cite{3381028,3439950}. 

\vspace{5pt}\noindent\textbf{Lifelong Learning}. 
Lifelong (or continual) learning refers to a system's ability to continually accommodate new knowledge to learn new tasks that were not predefined~\cite{978-3-642-79629-6-7}. Different approaches for lifelong learning exist based on supervised, unsupervised, and reinforcement learning~\cite{LML}, and recently based on neural networks~\cite{PARISI201954}. An interesting line of research is lifelong visual learning that combines computer vision with machine learning to enhance the understanding of visual scenes~\cite{LampertERC}. 
A key challenge for lifelong learning is dealing with catastrophic forgetting that refers to the loss of previous learning while learning new information; this may lead to failures for systems operating in real-world environments~\cite{HASSELMO2017407}. Approaches such as dynamic allocating new neurons or network layers to accommodate novel knowledge, or using complementary learning networks with experience replay do not yet provide the flexibility, robustness, and scalability required for real-world systems~\cite{PARISI201954}.

\vspace{5pt}\noindent\textbf{Why Lifelong Computing?}
Current approaches are not sufficient to tackle the growing demands on computing systems and the ever-changing conditions they face. The key underlying problem is that existing approaches lack an integrated perspective on handling change (anticipated and unanticipated) in an autonomous manner. While software evolution has been automated considerably in the past decades, in particular the deployment and integration of new computing elements, it remains in essence a human-driven activity. Autonomous and self-adaptive systems have expanded the operational domain of computing systems substantially, enabling them to deal with changes during operation, but the scope is in essence bounded to anticipated changes. Anomaly and novelty discovery mechanisms allow identifying deviations from expected behaviours, yet, these techniques are not ready to deal with high-dimensional multi-modal data, continuous data streams with concept drift, and complex anomalies. 

Lifelong learning enables learning algorithms dealing with new tasks that were not known at deployment time. Essentially, lifelong learning realises the evolution of a learning algorithm that learns from a continuous stream of data, without a predefined number of tasks to be learned.
Yet, besides the challenge of catastrophic forgetting, lifelong learning is tailored to learning new tasks only, which accounts just for one part of the overall problem of computing systems dealing with change autonomously. Emerging and future computing systems require the capabilities to self-adapt and self-evolve their own heterogeneous architecture that comprises computing and service elements, learning elements, and other resources. These capabilities will enable lifelong computing systems to change their architecture, including autonomous integration of new elements as needed, in response to the detection of anomalies, novelties, new goals or new constraints. This goes clearly far beyond the ability to deal with new learning tasks of lifelong learning. The integration and synergy between adaptation, evolution, and learning, three principle  approaches to deal with change, is the underlying foundation of lifelong computing but also one of its most demanding challenges. 

To tackle the challenges future computing systems face under the circumstances outlined here, a new holistic solution is required. Lifelong computing aims at offering such a solution.

\section{Future Offshore Aquaculture Scenario: Fish Farming}\label{sec:Aquaculture}  

We illustrate the need for lifelong computing systems with an example of a future offshore fish farm. The oceans, covering 71\% of the Earth’s surface, are generally considered as an important future source of food, minerals, and energy~\cite{DNV}. Yet, massive development of offshore infrastructures, such as wind farms and aquaculture farms, will put
increasing pressure on the oceans. It is therefore important to implement these infrastructures in a trustworthy and sustainable way to not threaten the marine ecosystems. 
We focus here on fish farming, which is traditionally done in coastal areas requiring human-intensive work with decisions based on farmers’ experience. For a future growth of fish production, there is a need to establish offshore farms and moving operations further away from the coast. This will make both logistics and operations become more complex and challenging.
Future offshore farms will consist of enclosed ocean areas, equipped with infrastructure for monitoring, feeding, cleaning, etc.~that need to be autonomously operated with a minimum or no human intervention. The realisation of offshore fish farms poses difficult conflicting challenges, including ensuring fish health and welfare, safety for people, fish, and structures, protecting the marine ecosystem, keeping aligned with new scientific insights as well as new technologies, economic viability with a sustainable and transparent supply chain, and climate-change resilience~\cite{EUaqua}. 
Tackling these challenges and balancing the trade-offs between the various system objectives will require an integrated computing system that is capable to operate, adapt and evolve autonomously throughout its lifetime in harsh and continuously changing environments. We illustrate how lifelong computing could offer such a unique solution.

\section{Blueprint Architecture for Lifelong Computing Systems}\label{sec:blueprint}

To deal with the challenges in managing their own adaptation and evolution throughout their lifetime, we put forward the following \textit{technical requirements} for lifelong computing systems: 
\begin{enumerate}
    \item To realise the goals of its users and handle a vast amount of data, lifelong computing systems should \textit{integrate different types of computing elements and learning algorithms};  
    \item A lifelong computing system should be able to \textit{adapt autonomously} satisfying and/or optimising multiple, potentially conflicting and dynamically changing objectives within ensured bounds, throughout its lifetime;
    \item A lifelong computing system should be able to \textit{discover and integrate new computing elements autonomously}; 
    \item To support its decision-making, a lifelong computing system should be able to \textit{exploit historical knowledge}, compliant with privacy and knowledge protection concerns; 
    \item A lifelong computing system should be self-aware and context-aware enabling it to self-learn, that is, \textit{autonomously evolving its architecture} to realise its goals within ensured bounds when anomalies, novelties, new goals or constraints are discovered, throughout its lifetime; 
    \item Depending on the domain at hand, some activities of a  lifelong computing system \textit{may be supported by humans}. 
\end{enumerate}

\begin{figure}[!thb]
    \centering
    \includegraphics[width=\linewidth]{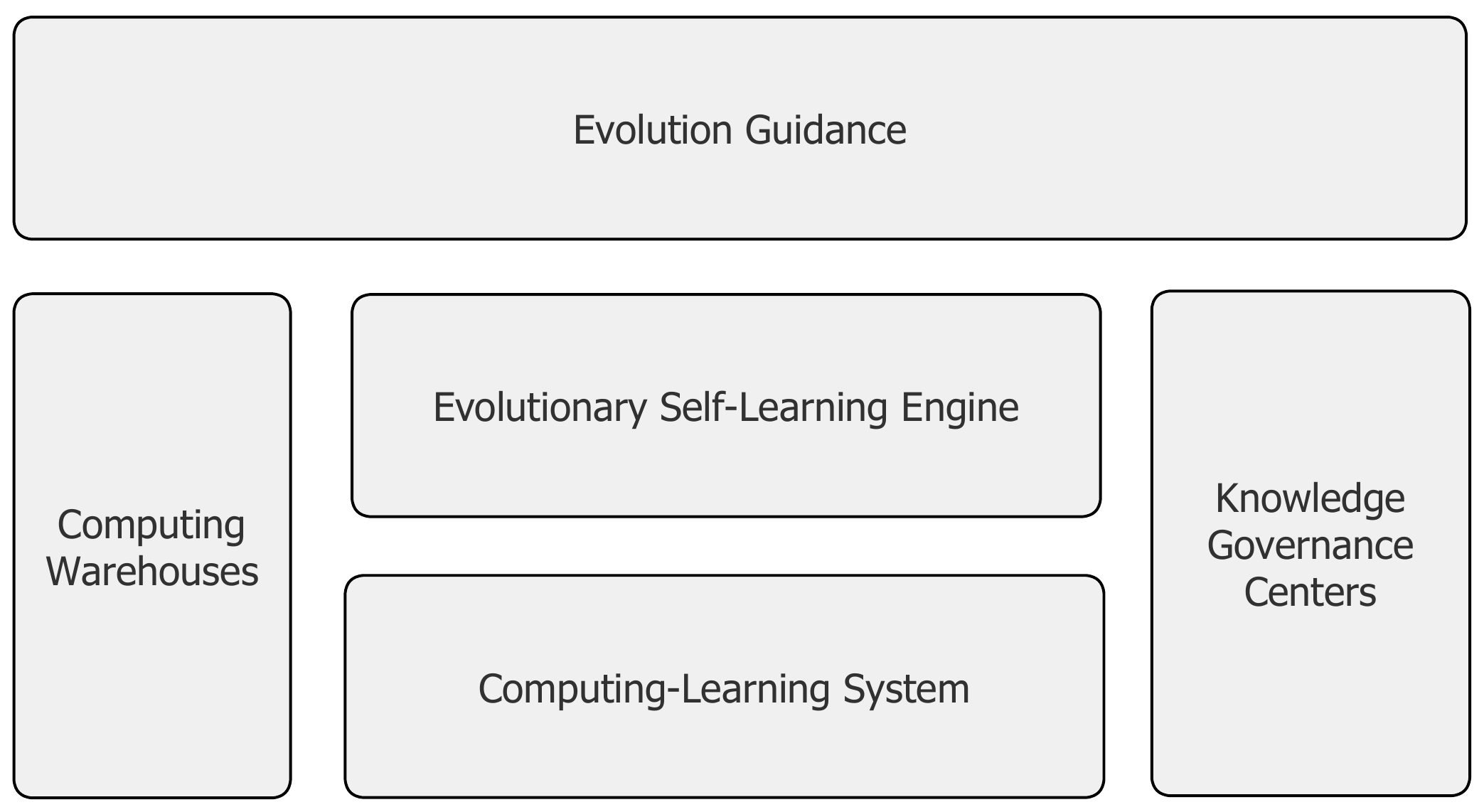}
    \caption{Blueprint architecture for lifelong computing systems with the different building blocks.}
    \label{fig:blueprint}
\end{figure}

To achieve these requirements, we propose a blueprint architecture for lifelong computing systems, shown in 
Figure~\ref{fig:blueprint}. We explain the different building blocks and illustrate each of them with examples of the offshore fish farming scenario. 

\vspace{5pt}\noindent\textbf{Computing-Learning System}. 
Central to lifelong computing is a  \textit{computing-learning system} that operates autonomously in a dynamic environment, realising requirements (1) and (2). To realise the user goals and process vast amounts of complex data, a  computing-learning system comprises a heterogeneous composition of  computing/service modules, learning modules, data stores, and other computing resources. 
To account for anticipated changes and balance the trade-offs between the various system goals, a  computing-learning system is equipped with a continuous 
and transparent 
multi-objective optimisation mechanism. This self-adaptation mechanism accounts for uncertainties and changing operation conditions that can be managed by parametric and structural changes of the running architectural configuration of the computing-learning system, without the need for updates of computing elements or the integration of new computing elements. To account for unanticipated changes (see evolutionary learning engine below), the computing-learning system should support automatic updates of its running architecture.

\begin{figure*}[!thb]
    \centering
    \includegraphics[width=\linewidth]{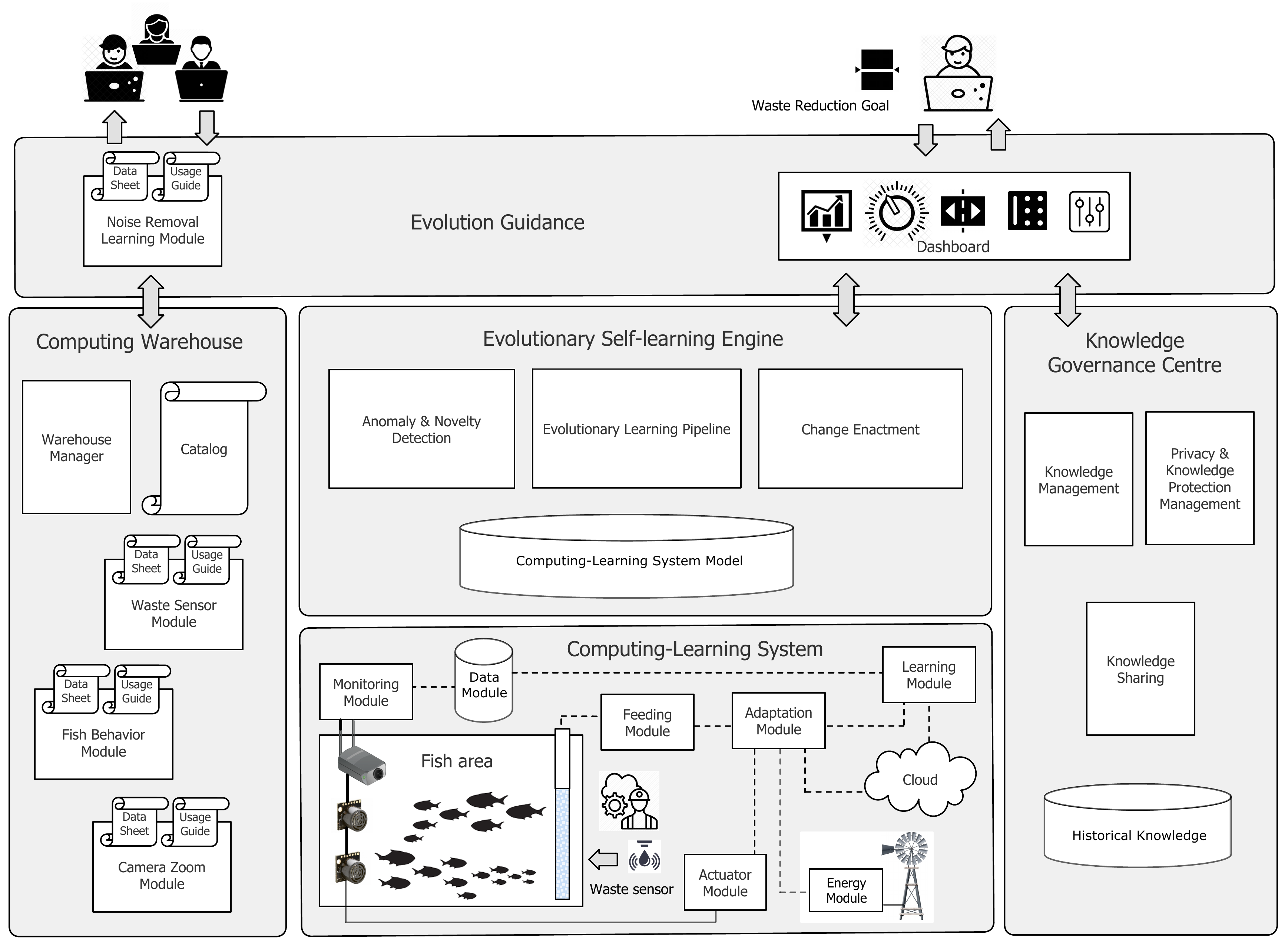}
    \caption{Illustration of blueprint architecture for offshore fish farm}
    \label{fig:blueprint-detail}
\end{figure*}

Figure~\ref{fig:blueprint-detail} illustrates a lifelong computing system for an offshore fish farm. 
We focus here on the computing-learning system (lower middle box) that comprises the fish area that is equipped with a variety of sensors (vision, hydro-acoustic sensors, etc.) that measure motion and behaviour of fish, health conditions of fish (e.g., wounds), food waste, temperature and oxygen level of the water, etc. The data is collected by a monitoring module and stored in a data repository. A learning module takes the data and 
learns and predicts relevant system healthiness indicators, such as fish health and well-being, biomass (size/age distributions), environmental footprint, etc.
These parameters together with other data obtained from local energy installations and the Cloud (e.g., weather forecasts) are then used by the adaptation module that continuously manages and optimises the multiple objectives of the fish farm and their trade-offs, using the feeding and actuator modules. For instance, when the oxygen level or PH-value of the water changes and more fish with wounds are detected, the frequency of measurements of particular sensors may be increased to provide higher temporal resolution to understand the problem and take action to resolve it.  

\vspace{5pt}\noindent\textbf{Computing Warehouses}. 
Lifelong computing systems are supported by  \textit{computing warehouses} that offer new computing elements realising requirement (3). These warehouses leverage the principles of off-the-shelf components and services, and open source software. Computing warehouses can be operated 
by software companies or by a broker. Example elements are 
software of a new camera (computing module), 
a connector to a new weather forecast service (service module), a template of new learning algorithm (learning module), etc. Crucial is that running lifelong computing systems can incorporate these computing elements  autonomously. To that end, each computing element is equipped with a \textit{data sheet} that specifies its functions, properties, usage requirements, etc., and a \textit{usage guide} that specifies procedures for using the element. All interactions with the computing warehouse happen via a \textit{warehouse manager}. Providers can offer new computing elements, possibly after certification. Clients can explore the available elements via a \textit{catalog} that lists the elements with links to the data sheets and usage guides. Using an element may be regulated by a contract. 

Figure~\ref{fig:blueprint-detail} shows a few examples of new elements for the fish farm (box left). The \textit{camera zoom module} provides the software that is required to activate and use a zoom lens on cameras. The \textit{fish behaviour module} may offer new  biological models of fish. The \textit{waste sensor module} offers the software to start using sensors that measure the waste of food at the ocean floor in the fish area. 

\vspace{5pt}\noindent\textbf{Knowledge Governance Centres}. Lifelong computing systems are supported by \textit{knowledge governance centres} that collect large amounts of data over time, process the data into reusable knowledge, and manage the knowledge compliant with privacy and knowledge protection concerns, realising requirement (4).\footnote{Knowledge identifies information about general concepts, data is information about specific instances; \url{https://link.springer.com/chapter/10.1007/978-1-4612-4980-1_8}} Such knowledge enables lifelong computing systems to transfer knowledge over time, different settings, and possibly different lifelong computing systems. The knowledge can be used in future decision-making, e.g., results of online experiments (see evolutionary self-learning engine below), for transfer learning, to deal with the problem of catastrophic forgetting, etc.  
Central to a knowledge governance centre is a repository that stores \textit{historical knowledge}.
The responsibilities of add and exploit knowledge is divided in several tasks. New data generated by lifelong computing systems is collected, processed (filter/merge/...), and stored in the repository by \textit{data management}. \textit{Knowledge sharing} provides lifelong computing systems access to use the knowledge. Access to the knowledge repository is regulated by \textit{privacy and knowledge protection management} that is responsible for ensuring that the knowledge is used compliant with the access policies, regulations and user's privacy requests~\cite{GDPR}. 
Knowledge governance centres can be operated by private software companies that run lifelong computing systems or by third parties that offer access to knowledge that is shared among lifelong computing systems. 

Figure~\ref{fig:blueprint-detail} illustrates a knowledge governance centre for a fish farm (box right). The historical knowledge repository may be located on the farm (with fast but possibly limited storage capacity), remotely (slow and expensive access to huge storage capacity), or a mix of both. Example knowledge maintained by the repository are classes of configurations of the computing-learning system used under different environmental conditions along with typical sensor values, actuator settings, and health indicator values. Other examples are summaries of relevant configurations with experiment results derived by lifelong computing algorithms. Such knowledge can be used by the evolutionary self-learning engine to support evolution of the computing-learning system (see below). Access to the knowledge repository should be controlled such that only the entities that need access (systems and humans) have access. Authentication and authorisation mechanisms can be used for this purpose. Furthermore, a protection strategy should ensure that knowledge can be restored quickly after corruption or loss, e.g., using Cloud backup.

\vspace{5pt}\noindent\textbf{Evolutionary Self-learning Engine}. At the heart of a lifelong computing system there is an \textit{evolutionary self-learning engine} that autonomously evolves the computing-learning system to handle any changes that cannot be handled by the built-in learning and adaptation mechanisms, realising requirement (5). Such unanticipated changes can be triggered either by the adaptation mechanism of the computing-learning system, by more disruptive \textit{anomaly or novelty detection} events the system has to deal with, or by new goals or constraints added to the system. Upon encountering such a trigger, the evolutionary self-learning engine will start to evolve its internal \textit{model of the  computing-learning system}. This runtime model contains a representation of the architecture of the system along with its goals (self-awareness) and relevant parts of the environment (context-awareness). The evolution of the model is conducted by the \textit{evolutionary learning pipeline}, which can be thought of as an evolutionary multi-objective learner that evolves the architecture of the computing-learning system, thereby exploiting new elements of computing warehouses and knowledge sources of knowledge governance centres.  
Using suitable methods and metrics for performance assessment of the evolving architectural model of the computing-learning system, the engine will evolve and optimise the computing-learning system model, thereby resulting in a novel architecture.
\textit{Change enactment} will then replace the running architecture of the computing-learning system with the novel architecture.  

As an example, assume that anomaly detection in Figure~\ref{fig:blueprint-detail} (middle box) discovers that the lenses of cameras are dirty resulting in poor quality images. To deal with this problem, a \textit{noise removal learning module} is added to the computing warehouse that offers a new learning algorithm, for instance a convolutional neural network to handle noisy images. To that end, the evolutionary self-learning engine runs online experiments evolving the model of the current architecture of the computing-learning system configuring and integrating the new noise removal learning module. The engine will use the resolution of images and improvements of healthiness indicators as performance metrics. In this process, the engine may exploit historical knowledge from the knowledge governance centre to accelerate the evolution process. Afterwards, the engine may store experimental results for later usage. Once the novel architecture is identified that satisfies the system goals, the current configuration will be evolved through change enactment. 

As another example, consider the need to  reduce food waste at the ocean floor in the fish area. To deal with this new goal, 
an operator adds a new \textit{waste reduction goal} to the evolutionary self-learning engine via the dashboard. 
The engine will then search in the catalog of the computing warehouse and find the waste sensor module. Based on the usage guidance provided by this module a set of \textit{new waste sensors} will be installed in the fish area by a field worker. The evolutionary self-learning pipeline will then evolve the architecture of the computing-learning system by extending the monitoring module with functionality to track waste and functionality to enable the actuator module to set and reconfigure the sensors (both derived from the waste sensor module). Furthermore, the new goal will be added to the multi-objective optimisation algorithm of the adaptation module. Finally, the learning module will be enhanced to take into account the data of the data module produced by the waste sensors. Once the new architecture is configured, it can be deployed via change enactment enabling the fish farm to reduce food waste by adjusting its settings, e.g. control the feeding,  based on observed conditions. 

\vspace{5pt}\noindent\textbf{Evolution Guidance}. Depending on the characteristics and requirements of the domain, a lifelong computing system may involve human experts to \textit{guide the evolution} of a lifelong computing system, realising requirement (6).  
Evolution guidance can range from a basic  dashboard that visualises key performance indicators of a lifelong computing system and provides ``knobs'' allowing operators to upload new computing elements and add new goals or constraints, up to full-fledged embodied AI that exploits intelligent user interface frameworks to  facilitate the evolution process of lifelong computing systems with the help of human interaction. 
Evolution guidance can support any of the functions of a lifelong computing system, for instance providing developers means to offer new computing elements (at computing warehouses), enabling operators to give feedback about discovered anomalies or novelties, adding a new goal or a new constraint (e.g., safety, privacy, energy consumption, environmental protection, or ethics), or give advice on architecture evolution (at the evolutionary self-learning engine), managing (access to) historical knowledge (at knowledge governance centres), among others. 

As an example for the fish farm, see  Figure~\ref{fig:blueprint-detail} (box at the top), evolution guidance enables  software developers to add a new learning module for noise removal to the computing warehouse. Furthermore, evolution guidance provides an interactive dashboard enabling an operator to support the evolutionary self-learning engine with identifying new software architectures of the computing-learning system. For instance, the operator may suggest (possibly new) quantitative and qualitative criteria (goals) to guide a multi-objective evolutionary pipeline in identifying new architectural configurations. The feedback of the operator may be incorporated into the fitness function allowing the learning pipeline to distinguish between promising and poor architectural configurations when evolving the model of the computing-learning system. A first approach towards involving a human expert in some evolution tasks was proposed in~\cite{RAMIREZ201892}. 

\section{Summary and Challenges Ahead}\label{sec:Challenges}

To conclude, we summarise the novelty of lifelong computing and we highlight key challenges to realise the newly proposed paradigm.

\small
\begin{table*}[t!]
\centering
\caption{Summary overview of the novelty of lifelong computing compared to the state of affairs}\label{tab:summary}
\begin{tabular}{m{2.35cm}m{2.2cm}m{2.4cm}m{1.9cm}m{2.2cm}m{2.5cm}m{1.8cm}}
\hline\hline\noalign{\smallskip}
Approach & Primary target & Handling \newline anticipated change & Integrating \newline new elements & Exploiting \newline historical data & Handling \newline unanticipated change & Role of humans\\
\noalign{\smallskip}\hline\noalign{\smallskip}
Autonomous systems & computing systems & AI techniques to \newline achieve goals under uncertainty 
(runtime) & -------------------- & runtime knowledge \newline representation \newline  (autonomous) & \vspace{8pt}--------------------------- & collaborator \newline (optional) \\\hline
Self-adaptation & computing systems & feedback loop to \newline achieve goals under uncertainty
(runtime) & -------------------- & runtime models \newline (autonomously) & \vspace{8pt}--------------------------- & adaptation guidance (optional)\\\hline
Software evolution & computing systems & --------------------------
& CI/CD \newline (tool-supported) & analysis data \newline  (offline) & stakeholder-driven \newline engineering (offline) & human-centred process\\\hline
SBSE & computing systems & meta-heuristic search method to solve problem 
(offline/runtime) & -------------------- & data sets/streams \newline (offline/runtime) &  \vspace{8pt}--------------------------- & 
determining fitness functions and constraints\\\hline
Anomaly/novelty \newline detection & computing systems & detect outliers of operational domain (offline/runtime)  & -------------------- & data sets/streams \newline (offline/runtime) & \vspace{8pt}--------------------------- & labelling data \newline (optional)\\\hline
Lifelong learning & learning algorithms & -------------------------- & -------------------- & training data \newline (semi-autonomous) & accommodate new \newline knowledge to learn \newline  new tasks (runtime) & labelling data \newline (optional) \\
\hline
\hline
Lifelong computing & computing-learning systems & multi-objective optimisation under uncertainty 
(runtime) & computing warehouses (autonomous) & knowledge governance centres \newline (autonomous) & evolutionary \newline self-learning (runtime) & adaptation and evolution guidance (optional)\\
\noalign{\smallskip}\hline\hline
\end{tabular}
\end{table*}
\normalsize

Table~\ref{tab:summary} summarises the novelty of lifelong computing in light of the state of affairs. For each state of the art approach, we list the primary type of system targeted, the approach (if any) to deal with anticipated change, the approach (if any) for integrating new computing elements, the approach (if any) for exploiting historical data, the approach (if any) to deal with unanticipated change, and finally the role of humans (if any).

The table shows why lifelong computing is fundamentally different from any existing approach, and why it is more than the sum of its parts. While the \textit{primary target} of existing approaches is either computing systems or learning algorithms, lifelong computing targets the integration of the two, matching the nature of emerging and future computing-learning systems. 

Several existing approaches are capable to \textit{deal with anticipated change} within the operational domain of the type of target system.
 Lifelong computing extends this to the operational domain of computing-learning systems. 

Software evolution is the only existing approach that supports the \textit{integration of new computing elements} in running systems using tool-supported CI/CD.
Lifelong computing systems on the other hand exploit computing warehouses, which allow them to autonomously select and integrate new computing elements during operation based on the needs at hand. 

Existing approaches \textit{exploit historical data}, yet, the data usage is limited to support the realisation of system functionality. 
Lifelong computing systems on the other hand exploit knowledge governance centres that provide the means to transfer knowledge over time and different settings. 

Software evolution \textit{handles unanticipated change} through a stakeholder-driven engineering process that updates or incorporates new software modules, while lifelong learning enables a learning algorithm to deal with new tasks during operation. Lifelong computing systems on the other hand are capable to handle unanticipated change autonomously through  evolutionary self-learning, incorporating new computing elements autonomously as needed. 

The \textit{role of humans} in existing approaches differ, ranging from being core entities that realise functionality as in software evolution and supervised lifelong learning to collaborators in autonomous systems. SBSE typically rely on humans to 
determine or adjust the fitness function and constraints, yet, the scope is limited to solving a particular problem using a heuristic search method. In contrast, lifelong computing systems act in essence completely autonomously to realise adaptation and evolution, throughout the system lifetime, supported by the provision of new computing elements. Optionally, humans can offer support in lifelong computing systems, for instance, for setting objectives or constraints on safety, privacy, etc., and providing guidance to support the evolutionary self-learning process.  

We have described how lifelong computing provides an answer to the lasting problem of how to engineer long running computing systems that can deal with a vast amount of highly complex data and autonomously adapt and evolve to deal with ever changing conditions, anticipated and unanticipated. Yet, realising the vision of lifelong computing, raises fundamental challenges. We list eight key achievements that are required to tackle these challenges: 

\begin{enumerate}
    \item A novel modelling approach for representing computing-learning systems that supports enacting safe online updates for 
    self-adaptation and system evolution;
    \item Novel continuous multi-objective optimisation mechanisms for runtime adaptation of computing-learning systems that operate under uncertain but anticipated changes;  
    \item The definition of standardised representations and interfaces of warehouse computing elements to facilitate their seamless integration into a computing-learning system's architecture;
    \item A novel approach to manage large amounts of historical data and turn this data into fruitful knowledge to support the autonomous evolution of computing-learning systems throughout their lifetime; 
    \item A novel family of unsupervised learning methods for discovering structure in complex, nonlinear, and high-dimensional data, such as images, videos, acoustic and other data (e.g. physiological data), which may be corrupted by noise, outliers and missing entries;
    \item A novel evolutionary self-learning pipeline for evolving computing-learning systems to deal with unanticipated changes; including multi-objective performance indicators for the estimation of variations of the computing-learning system by running them in an experimental sandbox;
    \item Novel mechanisms for guiding computing-learning systems' evolution, including notations and mechanisms for adding new goals or constraints, mechanisms for efficient information extraction, for context-based interpretation, and for explainability of decisions; 
    \item Novel mechanisms for providing guarantees (bounds) for multi-objective optimisation of computing-learning systems, for anomaly and novelty detection, and for evolutionary self-learning.
\end{enumerate}

Addressing these challenges requires a concerted research effort of experts from different specialisations within computer science: dynamic software architectures and dynamic learning architectures (to deal with the challenges of hybrid computing-learning systems), scalable multi-objective optimisation and self-adaptation (to deal with the challenges of adaptation of computing-learning systems), unsupervised learning for anomaly and novelty discovery, runtime models and self-awareness, and evolutionary learning (to deal with the challenges of evolutionary self-learning), 
and software engineering, data analysis/management (to deal with the challenges of computing warehouses, knowledge governance centres, and evolution guidance). Only the synergy between these specialisations, achieved through tight cooperation, can adequately yield solutions towards making the vision of lifelong computing come true.


\newlength{\bibitemsep}\setlength{\bibitemsep}{.2\baselineskip plus .05\baselineskip minus .05\baselineskip}
\newlength{\bibparskip}\setlength{\bibparskip}{0pt}
\let\oldthebibliography\thebibliography
\renewcommand\thebibliography[1]{%
  \oldthebibliography{#1}%
  \setlength{\parskip}{\bibitemsep}%
  \setlength{\itemsep}{\bibparskip}%
}


\balance
\bibliographystyle{ACM-Reference-Format}
\bibliography{main}


\begin{thebibliography}{41}


\ifx \showCODEN    \undefined \def \showCODEN     #1{\unskip}     \fi
\ifx \showDOI      \undefined \def \showDOI       #1{#1}\fi
\ifx \showISBNx    \undefined \def \showISBNx     #1{\unskip}     \fi
\ifx \showISBNxiii \undefined \def \showISBNxiii  #1{\unskip}     \fi
\ifx \showISSN     \undefined \def \showISSN      #1{\unskip}     \fi
\ifx \showLCCN     \undefined \def \showLCCN      #1{\unskip}     \fi
\ifx \shownote     \undefined \def \shownote      #1{#1}          \fi
\ifx \showarticletitle \undefined \def \showarticletitle #1{#1}   \fi
\ifx \showURL      \undefined \def \showURL       {\relax}        \fi
\providecommand\bibfield[2]{#2}
\providecommand\bibinfo[2]{#2}
\providecommand\natexlab[1]{#1}
\providecommand\showeprint[2][]{arXiv:#2}

\bibitem[\protect\citeauthoryear{B{\"a}ck}{B{\"a}ck}{1998}]%
        {Baeck1998}
\bibfield{author}{\bibinfo{person}{T. B{\"a}ck}.}
  \bibinfo{year}{1998}\natexlab{}.
\newblock \showarticletitle{An Overview of Parameter Control Methods by
  Self-Adaptation in Evolutionary Algorithms}.
\newblock \bibinfo{journal}{\emph{Fundamenta Informaticae}}
  \bibinfo{volume}{35}, \bibinfo{number}{1-4} (\bibinfo{year}{1998}),
  \bibinfo{pages}{51--66}.
\newblock


\bibitem[\protect\citeauthoryear{{Blair}, {Bencomo}, and {France}}{{Blair}
  et~al\mbox{.}}{2009}]%
        {Blair2009}
\bibfield{author}{\bibinfo{person}{G. {Blair}}, \bibinfo{person}{N. {Bencomo}},
  {and} \bibinfo{person}{R.~B. {France}}.} \bibinfo{year}{2009}\natexlab{}.
\newblock \showarticletitle{Models@ run.time}.
\newblock \bibinfo{journal}{\emph{Computer}} \bibinfo{volume}{42},
  \bibinfo{number}{10} (\bibinfo{year}{2009}), \bibinfo{pages}{22--27}.
\newblock
\urldef\tempurl%
\url{https://doi.org/10.1109/MC.2009.326}
\showDOI{\tempurl}


\bibitem[\protect\citeauthoryear{Boukerche, Zheng, and Alfandi}{Boukerche
  et~al\mbox{.}}{2020}]%
        {3381028}
\bibfield{author}{\bibinfo{person}{A. Boukerche}, \bibinfo{person}{L. Zheng},
  {and} \bibinfo{person}{O. Alfandi}.} \bibinfo{year}{2020}\natexlab{}.
\newblock \showarticletitle{Outlier Detection: Methods, Models, and
  Classification}.
\newblock \bibinfo{journal}{\emph{ACM Comput. Surv.}} \bibinfo{volume}{53},
  \bibinfo{number}{3}, Article \bibinfo{articleno}{55} (\bibinfo{date}{June}
  \bibinfo{year}{2020}), \bibinfo{numpages}{37}~pages.
\newblock
\showISSN{0360-0300}
\urldef\tempurl%
\url{https://doi.org/10.1145/3381028}
\showDOI{\tempurl}


\bibitem[\protect\citeauthoryear{Calinescu, Weyns, Gerasimou, Iftikhar, Habli,
  and Kelly}{Calinescu et~al\mbox{.}}{2018}]%
        {8008800}
\bibfield{author}{\bibinfo{person}{R. Calinescu}, \bibinfo{person}{D. Weyns},
  \bibinfo{person}{S. Gerasimou}, \bibinfo{person}{M.~U. Iftikhar},
  \bibinfo{person}{I. Habli}, {and} \bibinfo{person}{T. Kelly}.}
  \bibinfo{year}{2018}\natexlab{}.
\newblock \showarticletitle{Engineering Trustworthy Self-Adaptive Software with
  Dynamic Assurance Cases}.
\newblock \bibinfo{journal}{\emph{IEEE Transactions on Software Engineering}}
  \bibinfo{volume}{44}, \bibinfo{number}{11} (\bibinfo{year}{2018}),
  \bibinfo{pages}{1039--1069}.
\newblock
\urldef\tempurl%
\url{https://doi.org/10.1109/TSE.2017.2738640}
\showDOI{\tempurl}


\bibitem[\protect\citeauthoryear{Chen and Liu}{Chen and Liu}{2018}]%
        {LML}
\bibfield{author}{\bibinfo{person}{Z. Chen} {and} \bibinfo{person}{B. Liu}.}
  \bibinfo{year}{2018}\natexlab{}.
\newblock \bibinfo{booktitle}{\emph{Lifelong Machine Learning}}.
\newblock \bibinfo{publisher}{Morgan \& Claypool}.
\newblock
\showISBNx{9781681733029}


\bibitem[\protect\citeauthoryear{Cheng et~al\mbox{.}}{Cheng
  et~al\mbox{.}}{2009}]%
        {Cheng2009}
\bibfield{author}{\bibinfo{person}{B. Cheng} {et~al\mbox{.}}}
  \bibinfo{year}{2009}\natexlab{}.
\newblock \bibinfo{booktitle}{\emph{Software Engineering for Self-Adaptive
  Systems: A Research Roadmap}}.
\newblock \bibinfo{publisher}{Springer Berlin Heidelberg},
  \bibinfo{address}{Berlin, Heidelberg}, \bibinfo{pages}{1--26}.
\newblock
\showISBNx{978-3-642-02161-9}
\urldef\tempurl%
\url{https://doi.org/10.1007/978-3-642-02161-9_1}
\showDOI{\tempurl}


\bibitem[\protect\citeauthoryear{Commission}{Commission}{2021}]%
        {EU}
\bibfield{author}{\bibinfo{person}{European Commission}.}
  \bibinfo{year}{7/2021}\natexlab{}.
\newblock \showarticletitle{Advanced Computing}.
\newblock  (\bibinfo{year}{7/2021}).
\newblock
\urldef\tempurl%
\url{https://www.nsf.gov/funding/pgm_summ.jsp?pims_id=503306}
\showURL{%
\tempurl}


\bibitem[\protect\citeauthoryear{Dingsøyr, Nerur, Balijepally, and
  Moe}{Dingsøyr et~al\mbox{.}}{2012}]%
        {DINGSOYR20121213}
\bibfield{author}{\bibinfo{person}{T. Dingsøyr}, \bibinfo{person}{S. Nerur},
  \bibinfo{person}{V. Balijepally}, {and} \bibinfo{person}{N. Moe}.}
  \bibinfo{year}{2012}\natexlab{}.
\newblock \showarticletitle{A decade of agile methodologies: Towards explaining
  agile software development}.
\newblock \bibinfo{journal}{\emph{Journal of Systems and Software}}
  \bibinfo{volume}{85}, \bibinfo{number}{6} (\bibinfo{year}{2012}),
  \bibinfo{pages}{1213--1221}.
\newblock
\showISSN{0164-1212}
\urldef\tempurl%
\url{https://doi.org/10.1016/j.jss.2012.02.033}
\showDOI{\tempurl}
\newblock
\shownote{Special Issue: Agile Development.}


\bibitem[\protect\citeauthoryear{Elhabbash, Salama, Bahsoon, and
  Tino}{Elhabbash et~al\mbox{.}}{2019}]%
        {10.1145-3347269}
\bibfield{author}{\bibinfo{person}{A. Elhabbash}, \bibinfo{person}{M. Salama},
  \bibinfo{person}{R. Bahsoon}, {and} \bibinfo{person}{P. Tino}.}
  \bibinfo{year}{2019}\natexlab{}.
\newblock \showarticletitle{Self-Awareness in Software Engineering: A
  Systematic Literature Review}.
\newblock \bibinfo{journal}{\emph{ACM Trans. Auton. Adapt. Syst.}}
  \bibinfo{volume}{14}, \bibinfo{number}{2}, Article \bibinfo{articleno}{5}
  (\bibinfo{date}{Oct.} \bibinfo{year}{2019}), \bibinfo{numpages}{42}~pages.
\newblock
\showISSN{1556-4665}
\urldef\tempurl%
\url{https://doi.org/10.1145/3347269}
\showDOI{\tempurl}


\bibitem[\protect\citeauthoryear{Foundation}{Foundation}{2021}]%
        {NSF}
\bibfield{author}{\bibinfo{person}{National~Science Foundation}.}
  \bibinfo{year}{7/2021}\natexlab{}.
\newblock \showarticletitle{Computer Systems Research}.
\newblock  (\bibinfo{year}{7/2021}).
\newblock
\urldef\tempurl%
\url{https://www.nsf.gov/funding/pgm_summ.jsp?pims_id=503306}
\showURL{%
\tempurl}


\bibitem[\protect\citeauthoryear{Garlan, Cheng, Huang, Schmerl, and
  Steenkiste}{Garlan et~al\mbox{.}}{2004}]%
        {Rainbow}
\bibfield{author}{\bibinfo{person}{D. Garlan}, \bibinfo{person}{S. Cheng},
  \bibinfo{person}{A. Huang}, \bibinfo{person}{B. Schmerl}, {and}
  \bibinfo{person}{P. Steenkiste}.} \bibinfo{year}{2004}\natexlab{}.
\newblock \showarticletitle{Rainbow: Architecture-Based Self-Adaptation with
  Reusable Infrastructure}.
\newblock \bibinfo{journal}{\emph{Computer}} \bibinfo{volume}{37},
  \bibinfo{number}{10} (\bibinfo{date}{Oct.} \bibinfo{year}{2004}),
  \bibinfo{pages}{46–54}.
\newblock
\showISSN{0018-9162}
\urldef\tempurl%
\url{https://doi.org/10.1109/MC.2004.175}
\showDOI{\tempurl}


\bibitem[\protect\citeauthoryear{Grubbs}{Grubbs}{1969}]%
        {1969Grubbs}
\bibfield{author}{\bibinfo{person}{F.~E. Grubbs}.}
  \bibinfo{year}{1969}\natexlab{}.
\newblock \showarticletitle{Procedures for detecting outlying observations in
  samples}.
\newblock \bibinfo{journal}{\emph{Technometrics}} \bibinfo{volume}{11},
  \bibinfo{number}{1} (\bibinfo{year}{1969}).
\newblock


\bibitem[\protect\citeauthoryear{Harman, Burke, Clark, and Yao}{Harman
  et~al\mbox{.}}{2012}]%
        {6475391}
\bibfield{author}{\bibinfo{person}{M. Harman}, \bibinfo{person}{E. Burke},
  \bibinfo{person}{J.~A. Clark}, {and} \bibinfo{person}{X. Yao}.}
  \bibinfo{year}{2012}\natexlab{}.
\newblock \showarticletitle{Dynamic adaptive Search Based Software
  Engineering}. In \bibinfo{booktitle}{\emph{ACM-IEEE International Symposium
  on Empirical Software Engineering and Measurement}}. \bibinfo{pages}{1--8}.
\newblock
\urldef\tempurl%
\url{https://doi.org/10.1145/2372251.2372253}
\showDOI{\tempurl}


\bibitem[\protect\citeauthoryear{Harman, Jia, and Zhang}{Harman
  et~al\mbox{.}}{2015}]%
        {7102580}
\bibfield{author}{\bibinfo{person}{M. Harman}, \bibinfo{person}{Y. Jia}, {and}
  \bibinfo{person}{Y. Zhang}.} \bibinfo{year}{2015}\natexlab{}.
\newblock \showarticletitle{Achievements, Open Problems and Challenges for
  Search Based Software Testing}. In \bibinfo{booktitle}{\emph{2015 IEEE 8th
  International Conference on Software Testing, Verification and Validation
  (ICST)}}. \bibinfo{pages}{1--12}.
\newblock
\urldef\tempurl%
\url{https://doi.org/10.1109/ICST.2015.7102580}
\showDOI{\tempurl}


\bibitem[\protect\citeauthoryear{Harman and Jones}{Harman and Jones}{2001}]%
        {Harman2001}
\bibfield{author}{\bibinfo{person}{M. Harman} {and} \bibinfo{person}{B.~F.
  Jones}.} \bibinfo{year}{2001}\natexlab{}.
\newblock \showarticletitle{Search-based software engineering}.
\newblock \bibinfo{journal}{\emph{Information and Software Technology}}
  \bibinfo{volume}{43}, \bibinfo{number}{14} (\bibinfo{year}{2001}),
  \bibinfo{pages}{833--839}.
\newblock


\bibitem[\protect\citeauthoryear{Hasselmo}{Hasselmo}{2017}]%
        {HASSELMO2017407}
\bibfield{author}{\bibinfo{person}{M.~E. Hasselmo}.}
  \bibinfo{year}{2017}\natexlab{}.
\newblock \showarticletitle{Avoiding Catastrophic Forgetting}.
\newblock \bibinfo{journal}{\emph{Trends in Cognitive Sciences}}
  \bibinfo{volume}{21}, \bibinfo{number}{6} (\bibinfo{year}{2017}),
  \bibinfo{pages}{407--408}.
\newblock
\showISSN{1364-6613}
\urldef\tempurl%
\url{https://doi.org/10.1016/j.tics.2017.04.001}
\showDOI{\tempurl}


\bibitem[\protect\citeauthoryear{J{\"a}rvinen, Huomo, Mikkonen, and
  Tyrv{\"a}inen}{J{\"a}rvinen et~al\mbox{.}}{2014}]%
        {978-3-319-08738}
\bibfield{author}{\bibinfo{person}{J. J{\"a}rvinen}, \bibinfo{person}{T.
  Huomo}, \bibinfo{person}{T. Mikkonen}, {and} \bibinfo{person}{P.
  Tyrv{\"a}inen}.} \bibinfo{year}{2014}\natexlab{}.
\newblock \showarticletitle{From Agile Software Development to Mercury
  Business}. In \bibinfo{booktitle}{\emph{Software Business. Towards Continuous
  Value Delivery}}. \bibinfo{publisher}{Springer}.
\newblock


\bibitem[\protect\citeauthoryear{{Kephart} and {Chess}}{{Kephart} and
  {Chess}}{2003}]%
        {Kephart}
\bibfield{author}{\bibinfo{person}{J. {Kephart}} {and} \bibinfo{person}{D.
  {Chess}}.} \bibinfo{year}{2003}\natexlab{}.
\newblock \showarticletitle{The vision of autonomic computing}.
\newblock \bibinfo{journal}{\emph{Computer}} \bibinfo{volume}{36},
  \bibinfo{number}{1} (\bibinfo{year}{2003}), \bibinfo{pages}{41--50}.
\newblock


\bibitem[\protect\citeauthoryear{Lampert}{Lampert}{2021}]%
        {LampertERC}
\bibfield{author}{\bibinfo{person}{C. Lampert}.}
  \bibinfo{year}{7/2021}\natexlab{}.
\newblock \showarticletitle{Lifelong Learning of Visual Scene Understanding}.
\newblock \bibinfo{journal}{\emph{L3ViSU - 308036 - ERC grant}}
  (\bibinfo{year}{7/2021}).
\newblock
\urldef\tempurl%
\url{https://pub.ist.ac.at/~chl/erc/index.html}
\showURL{%
\tempurl}


\bibitem[\protect\citeauthoryear{Le~Goues, Dewey-Vogt, Forrest, and
  Weimer}{Le~Goues et~al\mbox{.}}{2012}]%
        {6227211}
\bibfield{author}{\bibinfo{person}{C. Le~Goues}, \bibinfo{person}{M.
  Dewey-Vogt}, \bibinfo{person}{S. Forrest}, {and} \bibinfo{person}{W.
  Weimer}.} \bibinfo{year}{2012}\natexlab{}.
\newblock \showarticletitle{A systematic study of automated program repair:
  Fixing 55 out of 105 bugs for \$8 each}. In \bibinfo{booktitle}{\emph{2012
  34th International Conference on Software Engineering (ICSE)}}.
  \bibinfo{pages}{3--13}.
\newblock
\urldef\tempurl%
\url{https://doi.org/10.1109/ICSE.2012.6227211}
\showDOI{\tempurl}


\bibitem[\protect\citeauthoryear{Li, Harman, Wu, and Zhang}{Li
  et~al\mbox{.}}{2017}]%
        {7582553}
\bibfield{author}{\bibinfo{person}{L. Li}, \bibinfo{person}{M. Harman},
  \bibinfo{person}{F. Wu}, {and} \bibinfo{person}{Y. Zhang}.}
  \bibinfo{year}{2017}\natexlab{}.
\newblock \showarticletitle{The Value of Exact Analysis in Requirements
  Selection}.
\newblock \bibinfo{journal}{\emph{IEEE Transactions on Software Engineering}}
  \bibinfo{volume}{43}, \bibinfo{number}{6} (\bibinfo{year}{2017}),
  \bibinfo{pages}{580--596}.
\newblock
\urldef\tempurl%
\url{https://doi.org/10.1109/TSE.2016.2615100}
\showDOI{\tempurl}


\bibitem[\protect\citeauthoryear{Li, Avgeriou, and Liang}{Li
  et~al\mbox{.}}{2015}]%
        {LI2015193}
\bibfield{author}{\bibinfo{person}{Z. Li}, \bibinfo{person}{P. Avgeriou}, {and}
  \bibinfo{person}{P. Liang}.} \bibinfo{year}{2015}\natexlab{}.
\newblock \showarticletitle{A systematic mapping study on technical debt and
  its management}.
\newblock \bibinfo{journal}{\emph{Journal of Systems and Software}}
  \bibinfo{volume}{101} (\bibinfo{year}{2015}), \bibinfo{pages}{193--220}.
\newblock
\showISSN{0164-1212}
\urldef\tempurl%
\url{https://doi.org/10.1016/j.jss.2014.12.027}
\showDOI{\tempurl}


\bibitem[\protect\citeauthoryear{Matthews, Reinerman-Jones, Barber, Teo,
  Wohleber, Lin, and Panganiban}{Matthews et~al\mbox{.}}{2016}]%
        {7573335}
\bibfield{author}{\bibinfo{person}{G. Matthews}, \bibinfo{person}{L.E.
  Reinerman-Jones}, \bibinfo{person}{D.J. Barber}, \bibinfo{person}{G. Teo},
  \bibinfo{person}{R.W. Wohleber}, \bibinfo{person}{J. Lin}, {and}
  \bibinfo{person}{A.R. Panganiban}.} \bibinfo{year}{2016}\natexlab{}.
\newblock \showarticletitle{Resilient autonomous systems: Challenges and
  solutions}. In \bibinfo{booktitle}{\emph{2016 Resilience Week (RWS)}}.
  \bibinfo{pages}{208--213}.
\newblock
\urldef\tempurl%
\url{https://doi.org/10.1109/RWEEK.2016.7573335}
\showDOI{\tempurl}


\bibitem[\protect\citeauthoryear{Meyer}{Meyer}{2014}]%
        {6802994}
\bibfield{author}{\bibinfo{person}{M. Meyer}.} \bibinfo{year}{2014}\natexlab{}.
\newblock \showarticletitle{Continuous Integration and Its Tools}.
\newblock \bibinfo{journal}{\emph{IEEE Software}} \bibinfo{volume}{31},
  \bibinfo{number}{03} (\bibinfo{date}{may} \bibinfo{year}{2014}),
  \bibinfo{pages}{14--16}.
\newblock
\showISSN{1937-4194}
\urldef\tempurl%
\url{https://doi.org/10.1109/MS.2014.58}
\showDOI{\tempurl}


\bibitem[\protect\citeauthoryear{Meyer-Nieberg and Beyer}{Meyer-Nieberg and
  Beyer}{2007}]%
        {Meyer-Nieberg2007}
\bibfield{author}{\bibinfo{person}{S. Meyer-Nieberg} {and} \bibinfo{person}{H.
  Beyer}.} \bibinfo{year}{2007}\natexlab{}.
\newblock \bibinfo{booktitle}{\emph{Self-Adaptation in Evolutionary
  Algorithms}}.
\newblock \bibinfo{publisher}{Springer Berlin Heidelberg},
  \bibinfo{pages}{47--75}.
\newblock
\showISBNx{978-3-540-69432-8}
\urldef\tempurl%
\url{https://doi.org/10.1007/978-3-540-69432-8_3}
\showDOI{\tempurl}


\bibitem[\protect\citeauthoryear{Mishra and Otaiwi}{Mishra and Otaiwi}{2020}]%
        {MISHRA2020100308}
\bibfield{author}{\bibinfo{person}{A. Mishra} {and} \bibinfo{person}{Z.
  Otaiwi}.} \bibinfo{year}{2020}\natexlab{}.
\newblock \showarticletitle{DevOps and software quality: A systematic mapping}.
\newblock \bibinfo{journal}{\emph{Computer Science Review}}
  \bibinfo{volume}{38} (\bibinfo{year}{2020}), \bibinfo{pages}{100308}.
\newblock
\showISSN{1574-0137}
\urldef\tempurl%
\url{https://doi.org/10.1016/j.cosrev.2020.100308}
\showDOI{\tempurl}


\bibitem[\protect\citeauthoryear{Moreno et~al\mbox{.}}{Moreno
  et~al\mbox{.}}{2015}]%
        {2786805.2786853}
\bibfield{author}{\bibinfo{person}{G.~A. Moreno} {et~al\mbox{.}}}
  \bibinfo{year}{2015}\natexlab{}.
\newblock \showarticletitle{Proactive Self-Adaptation under Uncertainty: A
  Probabilistic Model Checking Approach}. In \bibinfo{booktitle}{\emph{10th
  Joint Meeting on Foundations of Software Engineering}}.
  \bibinfo{publisher}{ACM}, \bibinfo{pages}{1–12}.
\newblock
\showISBNx{9781450336758}
\urldef\tempurl%
\url{https://doi.org/10.1145/2786805.2786853}
\showDOI{\tempurl}


\bibitem[\protect\citeauthoryear{Musić and Hirche}{Musić and Hirche}{2016}]%
        {MUSIC201642}
\bibfield{author}{\bibinfo{person}{S. Musić} {and} \bibinfo{person}{S.
  Hirche}.} \bibinfo{year}{2016}\natexlab{}.
\newblock \showarticletitle{Classification of human-robot team interaction
  paradigms}.
\newblock \bibinfo{journal}{\emph{IFAC-PapersOnLine}} \bibinfo{volume}{49},
  \bibinfo{number}{32} (\bibinfo{year}{2016}), \bibinfo{pages}{42--47}.
\newblock
\showISSN{2405-8963}
\urldef\tempurl%
\url{https://doi.org/10.1016/j.ifacol.2016.12.187}
\showDOI{\tempurl}
\newblock
\shownote{Cyber-Physical \& Human-Systems.}


\bibitem[\protect\citeauthoryear{Ouni and Mkaouer}{Ouni and Mkaouer}{2021}]%
        {10.1145-3449726.3461425}
\bibfield{author}{\bibinfo{person}{A. Ouni} {and} \bibinfo{person}{M.~W.
  Mkaouer}.} \bibinfo{year}{2021}\natexlab{}.
\newblock \showarticletitle{Search Based Software Engineering: Challenges,
  Opportunities and Recent Applications}. In \bibinfo{booktitle}{\emph{Genetic
  and Evolutionary Computation Conference Companion}} (Lille, France).
  \bibinfo{publisher}{ACM}, \bibinfo{address}{New York, NY, USA},
  \bibinfo{pages}{1032–1063}.
\newblock
\showISBNx{9781450383516}
\urldef\tempurl%
\url{https://doi.org/10.1145/3449726.3461425}
\showDOI{\tempurl}


\bibitem[\protect\citeauthoryear{Pang, Shen, Cao, and Hengel}{Pang
  et~al\mbox{.}}{2021}]%
        {3439950}
\bibfield{author}{\bibinfo{person}{G. Pang}, \bibinfo{person}{C. Shen},
  \bibinfo{person}{L. Cao}, {and} \bibinfo{person}{A.~Van~Den Hengel}.}
  \bibinfo{year}{2021}\natexlab{}.
\newblock \showarticletitle{Deep Learning for Anomaly Detection: A Review}.
\newblock \bibinfo{journal}{\emph{ACM Comput. Surv.}} \bibinfo{volume}{54},
  \bibinfo{number}{2}, Article \bibinfo{articleno}{38} (\bibinfo{date}{March}
  \bibinfo{year}{2021}), \bibinfo{numpages}{38}~pages.
\newblock
\showISSN{0360-0300}
\urldef\tempurl%
\url{https://doi.org/10.1145/3439950}
\showDOI{\tempurl}


\bibitem[\protect\citeauthoryear{Parisi, Kemker, Part, Kanan, and
  Wermter}{Parisi et~al\mbox{.}}{2019}]%
        {PARISI201954}
\bibfield{author}{\bibinfo{person}{G.~I. Parisi}, \bibinfo{person}{R. Kemker},
  \bibinfo{person}{J.~L. Part}, \bibinfo{person}{C. Kanan}, {and}
  \bibinfo{person}{S. Wermter}.} \bibinfo{year}{2019}\natexlab{}.
\newblock \showarticletitle{Continual lifelong learning with neural networks: A
  review}.
\newblock \bibinfo{journal}{\emph{Neural Networks}}  \bibinfo{volume}{113}
  (\bibinfo{year}{2019}), \bibinfo{pages}{54--71}.
\newblock
\showISSN{0893-6080}
\urldef\tempurl%
\url{https://doi.org/10.1016/j.neunet.2019.01.012}
\showDOI{\tempurl}


\bibitem[\protect\citeauthoryear{Ramírez, Romero, and Ventura}{Ramírez
  et~al\mbox{.}}{2018}]%
        {RAMIREZ201892}
\bibfield{author}{\bibinfo{person}{A. Ramírez}, \bibinfo{person}{J.~R.
  Romero}, {and} \bibinfo{person}{S. Ventura}.}
  \bibinfo{year}{2018}\natexlab{}.
\newblock \showarticletitle{Interactive multi-objective evolutionary
  optimization of software architectures}.
\newblock \bibinfo{journal}{\emph{Information Sciences}}
  \bibinfo{volume}{463-464} (\bibinfo{year}{2018}), \bibinfo{pages}{92--109}.
\newblock
\showISSN{0020-0255}
\urldef\tempurl%
\url{https://doi.org/10.1016/j.ins.2018.06.034}
\showDOI{\tempurl}


\bibitem[\protect\citeauthoryear{Rodríguez et~al\mbox{.}}{Rodríguez
  et~al\mbox{.}}{2017}]%
        {RODRIGUEZ2017263}
\bibfield{author}{\bibinfo{person}{P. Rodríguez} {et~al\mbox{.}}}
  \bibinfo{year}{2017}\natexlab{}.
\newblock \showarticletitle{Continuous deployment of software intensive
  products and services: A systematic mapping study}.
\newblock \bibinfo{journal}{\emph{Journal of Systems and Software}}
  \bibinfo{volume}{123} (\bibinfo{year}{2017}), \bibinfo{pages}{263--291}.
\newblock
\showISSN{0164-1212}
\urldef\tempurl%
\url{https://doi.org/10.1016/j.jss.2015.12.015}
\showDOI{\tempurl}


\bibitem[\protect\citeauthoryear{Tamai}{Tamai}{2019}]%
        {Tamai2019}
\bibfield{author}{\bibinfo{person}{Tetsuo Tamai}.}
  \bibinfo{year}{2019}\natexlab{}.
\newblock \bibinfo{booktitle}{\emph{Key Software Engineering Paradigms and
  Modeling Methods}}.
\newblock \bibinfo{publisher}{Springer International Publishing},
  \bibinfo{address}{Cham}, \bibinfo{pages}{349--374}.
\newblock
\showISBNx{978-3-030-00262-6}
\urldef\tempurl%
\url{https://doi.org/10.1007/978-3-030-00262-6_9}
\showDOI{\tempurl}


\bibitem[\protect\citeauthoryear{Thrun and Mitchell}{Thrun and
  Mitchell}{1995}]%
        {978-3-642-79629-6-7}
\bibfield{author}{\bibinfo{person}{S. Thrun} {and} \bibinfo{person}{T.~M.
  Mitchell}.} \bibinfo{year}{1995}\natexlab{}.
\newblock \showarticletitle{Lifelong Robot Learning}. In
  \bibinfo{booktitle}{\emph{The Biology and Technology of Intelligent
  Autonomous Agents}}. \bibinfo{publisher}{Springer},
  \bibinfo{pages}{165--196}.
\newblock
\showISBNx{978-3-642-79629-6}


\bibitem[\protect\citeauthoryear{Tzafestas}{Tzafestas}{2012}]%
        {Tzafestas2012}
\bibfield{author}{\bibinfo{person}{S.G. Tzafestas}.}
  \bibinfo{year}{2012}\natexlab{}.
\newblock \bibinfo{booktitle}{\emph{Advances in intelligent autonomous
  systems}}.
\newblock \bibinfo{publisher}{Springer}.
\newblock
\showISBNx{978-94-010-6012-7}


\bibitem[\protect\citeauthoryear{Union}{Union}{2021a}]%
        {EUaqua}
\bibfield{author}{\bibinfo{person}{European Union}.}
  \bibinfo{year}{7/2021}\natexlab{a}.
\newblock \bibinfo{booktitle}{\emph{Aquaculture Policy}}.
\newblock \bibinfo{type}{{T}echnical {R}eport}.
\newblock
\urldef\tempurl%
\url{https://ec.europa.eu/oceans-and-fisheries/policy/aquaculture-policy_en}
\showURL{%
\tempurl}


\bibitem[\protect\citeauthoryear{Union}{Union}{2021b}]%
        {GDPR}
\bibfield{author}{\bibinfo{person}{European Union}.}
  \bibinfo{year}{7/2021}\natexlab{b}.
\newblock \bibinfo{booktitle}{\emph{Data protection in the EU}}.
\newblock \bibinfo{type}{{T}echnical {R}eport}.
\newblock
\urldef\tempurl%
\url{https://ec.europa.eu/info/law/law-topic/data-protection/data-protection-eu_en}
\showURL{%
\tempurl}


\bibitem[\protect\citeauthoryear{Veritas}{Veritas}{2020}]%
        {DNV}
\bibfield{author}{\bibinfo{person}{Det~Norske Veritas}.}
  \bibinfo{year}{2020}\natexlab{}.
\newblock \showarticletitle{Technology Outlook 2030 - Safer, Smater, Greener}.
\newblock  (\bibinfo{year}{2020}), \bibinfo{pages}{1--110}.
\newblock
\urldef\tempurl%
\url{www.dnvgl.com}
\showURL{%
\tempurl}


\bibitem[\protect\citeauthoryear{Weyns}{Weyns}{2020}]%
        {weyns2020book}
\bibfield{author}{\bibinfo{person}{D. Weyns}.} \bibinfo{year}{2020}\natexlab{}.
\newblock \showarticletitle{Introduction to Self-Adaptive Systems: A
  Contemporary Software Engineering Perspective}.
\newblock \bibinfo{publisher}{Wiley}.
\newblock
\newblock
\shownote{ISBN 978-1-119-57494-1.}


\bibitem[\protect\citeauthoryear{Wooldrige}{Wooldrige}{2009}]%
        {Wooldrige2009}
\bibfield{author}{\bibinfo{person}{M. Wooldrige}.}
  \bibinfo{year}{2009}\natexlab{}.
\newblock \showarticletitle{An Introduction to MultiAgent Systems}.
\newblock \bibinfo{publisher}{Wiley}.
\newblock
\newblock
\shownote{ISBN 978-0-470-51946-2.}


\end{thebibliography}

\end{document}